\documentstyle[prd,aps,floats,epsfig,times]{revtex}
\draft
\begin{document}
\twocolumn[\hsize\textwidth\columnwidth\hsize\csname@twocolumnfalse\endcsname

\title{Self interacting Brans Dicke cosmology and
Quintessence}
\author{S. Sen$^{\star}$ and T. R. Seshadri$^{\dagger}$}
\address{Mehta Research Institute, Chhatnag Road,
Jhusi. Allahabad 211
019 India}
\date{{\today}}
\maketitle
\begin{abstract}
Recent cosmological observations reveal that we are
living in a flat
accelerated expanding universe. In this work we have
investigated the
nature
of the potential compatible with the power law
expansion of the
universe in
 a self interacting Brans Dicke cosmology with a
perfect fluid
background and
have analyzed whether this potential supports the
accelerated expansion. It
is found
 that positive power law potential is relevant in this
scenario and can
drive
 accelerated expansion for negative Brans Dicke
coupling parameter
$\omega$.
The evolution of the density perturbation
is also analyzed in this scenario 
and is seen that the model allows
growing modes for
negative
$\omega$.
\end{abstract}
\pacs{PACS Number(s): 04.20Jb, 98.80Hw\hfill{\sc
MRI-PHY/P20000532}}
\vskip 2pc]

\section{Introduction}
It has been suggested by a number of recent
observations\cite{R1} that
merely the  baryonic matter and dark matter
 may not be able to account for the total
density of the universe. This suggests that
either the universe is open (density less than the
critical density), or, that
there is yet another source of energy density
which makes the universe spatially flat ($\Omega_{tot}
=1$). The anisotropy of the cosmic microwave
background radiation (CMBR) as indicated
by the recent results from Boomerang\cite{R2} strongly favour
the
second
possibility of a flat universe which makes the issue of
missing energy
more
important. Quintessence\cite{R3} has been proposed as
that missing
energy density
component that along with the matter and baryonic
density makes the
density parameter equal to $1$.
\par
 The luminosity-redshift relation observed
for the Type-Ia supernovae \cite{R5} strongly suggests
that, in the present phase,
the universe is undergoing an accelerated expansion.
This is supported also by the recent measurements of
CMBR and the power spectrum of mass perturbations
\cite{R4}.
 Of course, the  simplest and the most extreme
choice
in such a case, is  the model with the vacuum
energy density or the cosmological constant
$\Lambda$\cite{R5*}, but other  options also
exist.
 Models with quintessence and cold dark matter
(QCDM) is one interesting
possibility.
Basically, quintessence is a dynamical slowly evolving
spatially
inhomogeneous component of energy density with negative
pressure\cite{R3}.
The energy density associated with a scalar
field $Q$ slowly
 moving down its
potential $V$ can represent a simple example of
Quintessence\cite{R6}.
The potential energy of the scalar field should
dominate over the
kinetic
energy of the field for slow rolling and thereby
making the pressure
negative $(p_Q={1\over{2}}\dot Q^2-V)$. For
quintessence, the equation
of
state defined by $\gamma_Q(={p_Q\over{\rho_Q}})$ lies
between 0 and
$-1$.
 More precisely $-0.6\geq \gamma_Q \geq -1$.
The
deceleration parameter
$q$ $({\bf =}-{\ddot{R}R\over{\dot R^2}})$ accounting
for the acceleration of the
universe also lies in the same range 0 and $-1$.
Depending on the form
of $V(Q)$, $\gamma_Q$ can be constant, slowly varying,
rapidly varying
or
oscillatory\cite{R3}.
\par
This quintessence proposal faces two types of
problems\cite{R7}.
One of these problems,
 (referred as fine tuning problem), is the smallness of
the
energy density compared to other typical particle
physics scales. The
other
problem known as the cosmic coincidence problem,
 is that although
the missing energy density and matter density
decrease at different
rates as
the universe expands, it appears that the initial
condition has to be
set so
precisely that the two densities become comparable
today. A special
form of
quintessence field, called the `tracker field',
has been proposed to tackle
this problem\cite{R8}. The good point
about these models is that they have attractor like
solutions that make
the
present time behaviour nearly independent of the
initial conditions.
\par
 A scalar field with a potential dominating over
its kinetic
energy
constitutes the simplest example of
quintessence. Quite a few models have also been
suggested with a minimally coupled scalar field and
different types of
potentials. A purely exponential potential
is one of the widely
studied cases \cite{R9}. In spite of the
other advantages the energy density is not enough to make
up for the missing part.
Inverse power law is the other potential
(\cite{R6}-\cite{R8}) that has
been studied extensively for quintessence models,
particularly for
solving
the cosmic coincidence problem. Though the problems
are resolved
successfully
with this potential, the predicted value for
$\gamma_Q$ is not in good
agreement with the observed results. In search
of proper
potentials that
 would eliminate the problems,
new types of potentials, like $V_0[\cosh
\lambda\phi-1]^p$\cite{R10}
and
$V_0\sinh (\alpha\sqrt
k_0\Delta\phi)^\beta$\cite{R5*,R11} have been
considered,
which
 have
asymptotic forms like the inverse power law or
exponential ones.
 Different physical considerations have lead to
the study of other types of the potentials
also\cite{R13}.Recently Saini {\it et al}  \cite{Rsai} have 
reconstructed the potential 
in context of general relativity and minimally coupled quintessence 
field from the expression of the luminosity distance $d_L(z)$
as function of redshift obtained from the observational data.  
However, none of these are entirely free of
problems. Hence, there is still
a need to identify appropriate potentials to explain
current observations
\cite{R9}.
\par
As we have mentioned earlier most of the studies
to produce variable $\Lambda$ have been done with a 
minimally coupled scalar
field representing the quintessence field. 
 It has been recently shown by Pietro 
and Demaret\cite{Rpietro} that for constant scalar field equation of 
state, which is a good approximation for a tracker field solutions, 
the field equations and the conservation equations strongly constrain 
the scalar field potential, and most of the widely used potential for 
quintessence, such as inverse power law one, exponential or the cosine 
form, are incompatible with these constraints. The minimally coupled 
self interacting models will also be ruled out if the observations 
predict that the missing component of the energy density obeys 
an equation of state $p=\gamma\rho$ with $\gamma<-1 (\rho\geq 0)$ , 
and these sort of equation of state is in reasonable agreement with 
different observations \cite{Rcald}. These facts motivates toward a 
more general theory like scalar tensor theory. Scaling
attractor
solutions
are available in the literature with the
exponential\cite{R13} and
power law
\cite{R13,R14} potentials in non-minimally coupled
theories. There are different approach also to find
tracking solutions
 in general scalar tensor theories with
inverse power law
potentials\cite{R15}. Bertolami
et. al.\cite{R16} have found
accelerating solutions with
quadratic
potentials in Brans Dicke cosmology.
In a similar scenario Ritis et al \cite{R17}
found a family of tracking
solutions.  We have also found accelerating solutions 
in Brans Dicke Cosmology with a potential which has a time dependent
mass squared term which has recently become negative\cite{Rasen}.
Faraoni\cite{R18} have studied
different potentials with a non-minimal coupling term
$\psi R{\phi^2\over{2}}$.
Like Saini {\it et al}\cite{Rsai}, Boisseau {\it et al} \cite{Rboi} 
have reconstructed the 
potential from the luminosity-redshift relation available from the 
observations in context of scalar tensor theory.
\par
In this work we investigate the nature of the
potential relevant to the
power law expansion of the universe in a self
interacting Brans Dicke
Cosmology with a perfect fluid distribution. 
The Jordan Brans Dicke (JBD) action \cite{R19}
with a self interacting potential and matter is
represented by,
\begin{equation}
{\cal{S}}=\int d^4x \sqrt{-g}[\phi
R-{\omega\over{\phi}}\phi^\alpha\phi_\alpha
+{\cal{L}}_m]
\end{equation}
where $\omega$ is the Brans Dicke parameter and
${\cal{L}}_m$ is the
Lagrangian
of the matter field. In this theory ${1\over{\phi}}$
plays the role of
the
gravitational constant. This action also matches
with the low energy string
theory
action\cite{R20} for $\omega=-1$.
It is
checked whether the accelerated cosmic expansion can be
driven by this potential.
The density perturbation is also studied to check the
consistency of the
structure
formation scenario. In the next section we present the
exact solution
for the
field equations and investigate the nature of the
solution. Section 3 discusses
the density perturbation part
of the solution.
In the last section we have
discussed about the different features of the model.

%%%%%%%%%%%%%%%%%%%%%SECTION II
%%%%%%%%%%%%%%%%%%%%%%%%%%%%%%%%%%%%%%
\section{Field equations and solutions}
We start with the Brans Dicke action along with a self
interacting potential and a matter field
\begin{equation}
{\cal{S}}=\int d^4x \sqrt{-g}[\phi
R-{\omega\over{\phi}}\phi^\alpha\phi_\alpha
-V(\phi)+{\cal{L}}_m]
\end{equation}
(We have chosen the unit $8\pi G_0=c~=~1$.)

%%%%%%%%%%%%%%%%%%%%%%
%I thought one need not give equations 3 to 5 in your
%%%%%%%%%%%%%%%%%%%%%%%%%%%%%%%%%%%%
The matter content of the universe is composed of
perfect fluid
\begin{equation}
T_{\mu\nu}=(\rho+p){\it{v}}_\mu{\it{v}}_\nu+p
g_{\mu\nu},
\end{equation}
where ${\it{v}}_\mu{\it{v}}^\mu=-1$.
 We assume that the universe is homogeneous,
isotropic and spatially
flat. Such a universe is described by the FRW
line-element,
\begin{equation}
ds^2=-dt^2+R^2(t)[dr^2+r^2 d\theta^2+r^2\sin^2\theta
d\phi^2]
\end{equation}
 The Einstein's field equations and the evolution
equation for the scalar field
are given by,
\begin{equation}
3{\dot R^2\over{R^2}}+3{\dot
R\over{R}}{\dot\phi\over{\phi}}-{\omega\over{2}}{\dot\phi^2\over{\phi^2}}-{V\over{2\phi}}={\rho\over{\phi}},
\end{equation}
\begin{equation}
2{\ddot{R}\over{R}}+{\dot R^2\over{R^2}}+{\ddot
{\phi}\over{\phi}}+2{\dot
R\over{R}}{\dot\phi\over{\phi}}+{\omega\over{2}}{\dot\phi^2\over{\phi^2}}-{V\over{2\phi}}=-{p\over{\phi}}
\end{equation}
\begin{equation}
{\ddot{\phi}}+3{\dot
R\over{R}}{\dot\phi}={\rho-3p\over{2\omega+3}}-{1\over{2\omega+3}}\left[2V-\phi{dV\over{d\phi}}\right]
\end{equation}
The energy conservation equation, that follows from
the Bianchi
identity gives
\begin{equation}
\dot\rho+3{\dot R\over{R}}(\rho+p)=0
\end{equation}
Among these four equations only three are independent.
But as there are
five
unknowns $(R, \phi,\rho,p,V)$ two assumptions can be
made to match the
number
of unknowns with the number of independent equations.
With this freedom, we
 choose the functional form for the time-evolution of the scale factor
and
the scalar field
and
find the potential that is compatible with this choice. 
We consider solutions for which the scale factor and the field $\phi$
evolve as power-law functions of time.
\begin{equation}
R=R_0 \left({t\over{t_0}}\right)^\alpha~~~~{\rm{and}}~~~~
\phi=\phi_0 \left({t\over{t_0}}\right)^\beta \label{Rphi}
\end{equation}
 where the subscript $0$ refers to the values
of the parameters at the present epoch and $t_0$ is
the present epoch, i.e the
age of the universe.
\par
 In order to get a solution with accelerated
expansion,
the deceleration parameter, $q$, has to be negative.
This restricts the parameter $\alpha$
to be greater than $1$.
\par
The solutions for $\rho$ and $p$ can be found to be 
\begin{eqnarray}
\rho & = & \rho_c t^{\beta-2}  \label{rho=rhocetc}\\
p & = & p_c t^{\beta-2} \label{p=pcetc}
\end{eqnarray}
where,
\begin{eqnarray}
\rho_c & = & {3\alpha\phi_0\over{t_o^\beta}}
\left[{2\alpha+\beta(1+\alpha)-\beta^2(1+\omega)\over{2-\beta}}\right]
\label{rhocdef}\\
p_{c} & = & {(2-\beta-3\alpha)\phi_0\over{t_o^\beta}}
\left[{2\alpha+\beta(1+\alpha)-\beta^2(1+\omega)\over{2-\beta}}\right]
\label{pcdef}\\
\end{eqnarray}
The form of the potential allowed by this ansatz is
\begin{equation}
V = \bar{V_c} \phi^{\beta-2\over{\beta}}
\end{equation}
where,
\begin{equation}
\bar{V_c}=
{\beta\phi_0^{2\over{\beta}}\over{t_o^2}}{\left[6\alpha(\omega+1)-
\omega\beta(2-\beta)-12\alpha^2\right]\over{2-\beta}}
\end{equation}

From (\ref{rho=rhocetc}), (\ref{p=pcetc}),
(\ref{rhocdef}) and (\ref{pcdef}) it is
clear that the
background fluid follows an equation of state
of the form 
$p=\gamma_\beta \rho$
where $\gamma_\beta$ is given by
\begin{equation}
\gamma_B={2-\beta\over{3\alpha}}-1. \label{gammaB}
\end{equation}
Let us now see how the parameters are constrained.
First of all we are interested in a scenario in which the universe
undergoes an accelerated expansion. As we have mentioned earlier
this immediately
constrains $\alpha$ to be greater than $1$.
Further, $\gamma_B$, for perfect fluid,  is constrained to be between
$0$ and $1$,
From equation (\ref{gammaB}), this implies $2-6\alpha<\beta<2-3\alpha$.
We may note that since $\alpha>1$, $\beta$ is always negative.

To clearly specify the nature of the
expansion, and the missing energy we investigate
further the energy density and the pressure of
the Brans Dicke
field. From the field equations, the energy density
and the pressure of
the scalar field turns out to be
\begin{equation}
\rho_\phi=
\left[{\omega\over{2}}~{\dot\phi^2\over{\phi}}+{V\over{2}}-3{\dot
R\over{R}}\dot\phi\right]
\end{equation}
and
\begin{equation}
p_\phi=\left[{\omega\over{2}}~{\dot\phi^2\over{\phi}}-{V\over{2}}+\ddot\phi
+2{\dot R\over{R}}\dot\phi\right]
\end{equation}
The expression for the energy density and pressure of the Brans Dicke 
are respectively  given by,
\begin{equation}
\rho_\phi={3\alpha\beta\over{2-\beta}}~\{\beta(1+\omega)-2\alpha-1\}
{\phi_0\over{t_o^\beta}}t^{\beta-2} \label{rhophi}
\end{equation}
\begin{eqnarray}
p_\phi & = &
{\beta\over{2-\beta}}\{2\alpha+\beta(1+\alpha)-\beta^2(1+\omega)\nonumber\\
& ~~~~~+ &
(2-3\alpha)[\beta(1+\omega)-(2\alpha+1)]\}{\phi_0\over{t_o^\beta}}t^{\beta-2}
\label{pphi}
\end{eqnarray}

In order to ensure the positivity condition for both scalar field and matter, we find 
that $\omega$ is constrained as,
$$\frac{1+2\alpha}{\beta}-1<\omega<\frac{1+2\alpha}{\beta}-1+\frac{\alpha(2-\beta)}{\beta^2}.$$
The allowed region of $\omega$ in $(\beta,\omega)$ parameter space
is shown in figure 1 for $\alpha=1.1$.
\begin{figure}[h]
%\psrotatefirst
\centering
\leavevmode\epsfysize=8cm\epsfbox{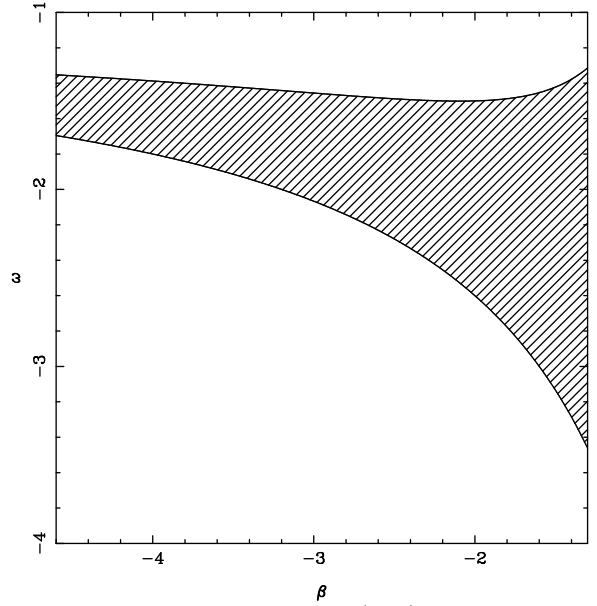}
\caption{The allowed region in the $(\beta,\omega)$
parameter space for $\alpha=1.1$.}
\label{figure1}
\end{figure}
Now from (\ref{rhophi}) and (\ref{pphi}), it is evident that the scalar
field describes an
equation of state $p_\phi=\gamma_\phi\rho_\phi$ where
the $\gamma_\phi$
is given by
\begin{equation}
\gamma_\phi={2\over{3\alpha}}-1+{2\alpha+\beta(1+\alpha)-\beta^2(1+\omega)\over{3\alpha\{\beta(1+\omega)-2\alpha-1\}}}
\end{equation}
The expressions for the density parameters for the matter and the 
scalar field, defined respectively by $\Omega_m=\frac{\rho}{3H^2\phi}$
 and $\Omega_\phi=\frac{\rho_\phi}{3H^2\phi}$\cite{Rdiaz}, are given by
\begin{equation}
\Omega_m=\frac{2\alpha+\beta(1+\alpha)-\beta^2(1+\omega)}{\alpha(2-\beta)}
\end{equation}
and
\begin{equation}
\Omega_\phi=\frac{\beta^2(1+\omega)-2\alpha\beta-\beta}{\alpha(2-\beta)}
\end{equation}
Using these two expressions $\gamma_\phi$ can be recast in terms of 
$\Omega_m$ and $\Omega_\phi$ as 
\begin{equation}
\gamma_\phi=\frac{1}{3\alpha}\left(2+\beta\frac{\Omega_m}{\Omega_\phi}\right)-1
\end{equation} 
As we have mentioned earlier that the result of the Supernova Cosmology 
Project and the High $z$ Survey Project that reveal the so-called 
acceleration of the universe, favours a universe with positive 
cosmological constant as well\cite{R5}. Assuming flatness in context of 
general relativity the best fit for these data occurred for $\Omega_m=0.28$
and $\Omega_\Lambda=0.72$. Using these best fit values along with the 
bound on $\beta$, it is found that $\gamma_\phi$ lies within the range
$$-0.87<\gamma_\phi<-0.47$$
 In the following two figures we find the allowed region (black)
 in the 
$(\alpha,\beta)$ 
parameter space which has a 
$\gamma_\phi~~
(-0.6~~{\rm and}~~-0.7$ with the matter density range within 
$0.2<\Omega_m<0.4$.
\par
In Brans Dicke theory, the value of the gravitation constant is determined 
by the value of $\phi$
\begin{equation}
G(t)={1\over{\phi(t)}},
\end{equation}
The scalar field $\phi$ approaches to constant value $\phi_0$ 
at the present time and consequently the inverse of $\phi_0$ 
gives the Newtonian constant $G_N$.
The rate of change of $G$ at present time is 
given by
$\left[{\dot G\over{G}}\right]_{t=t_0}= -{\beta\over{\alpha}}H_0$
where $H_0$ is the Hubble parameter at present time 
$(\equiv {\dot{R}\over{R}}(t_0)={\alpha\over{t_0}})$. 
For any specific value of
$\alpha$ and $\beta$ admitting accelerated expansion, we need to 
${\dot G\over{G}}<10^{-10}$ per year \cite{R25}.\\
\begin{figure}[hb]
\centering
\leavevmode\epsfysize=7cm \epsfbox{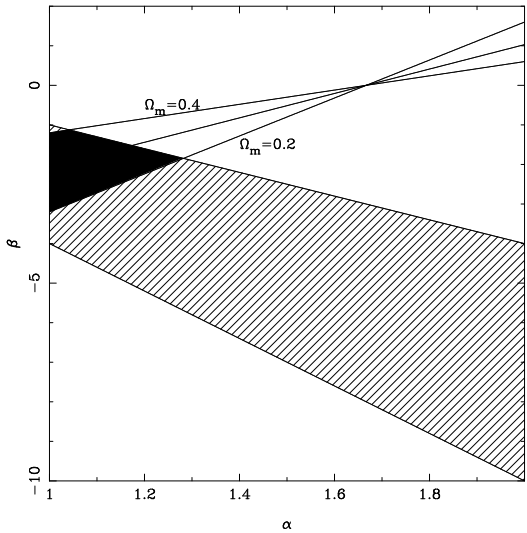}
\caption{$\beta$ {\it vs} $\alpha$ for different $\Omega_m$ values and $\gamma_\phi=-0.6$}
\label{figure2}
\end{figure}
\begin{figure}[hb]
\centering
\leavevmode\epsfysize=7cm \epsfbox{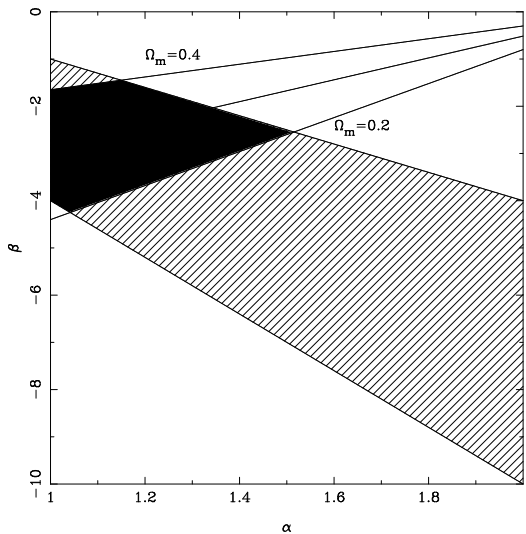}
\caption{$\beta$ {\it vs} $\alpha$ for different $\Omega_m$ values and $\gamma_\phi=-0.7$}
\label{figure3}
\end{figure}

In a recent investigation Pont et al\cite{Rpont} estimated the age 
of the universe $t_0$ to be $14\pm 2$ Gyrs, that means $\alpha$, 
which is given by $H_0t_0$,
 should be very close to 1.
 At the same time Kaplinghat et al\cite{Rkapling} and others\cite{Rsethi}
 have also pointed out that
for power law cosmologies high redshift data indicates $\alpha$ to be 
$\approx 1$. In fact in one of our recent investigation we have shown that 
the best fit value for $\alpha$ with SNIa data for power law cosmology is 
approximately 1.25\cite{Rsen}. 
But $H_0t_0$ is also determined by the expression
\begin{equation}
(H_0t_0)^2=\left[{\alpha
\{2\alpha+\beta(1+\alpha)-\beta^2(1+\omega)\}\over{\Omega_M
(2-\beta)}}\right] \label{H0t0}
\end{equation}\\
\begin{figure}[hb]
\centering
\leavevmode\epsfysize=7cm \epsfbox{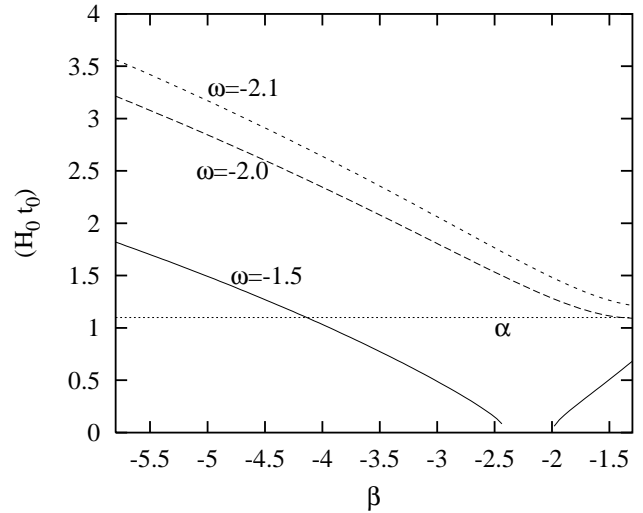}
\caption{$H_0 t_0$ {\it vs} $\beta$ for different $\omega$ values and $\alpha=1.1$}
\label{figure4}
\end{figure}
\begin{figure}[hb]
\centering
\leavevmode\epsfysize=7cm \epsfbox{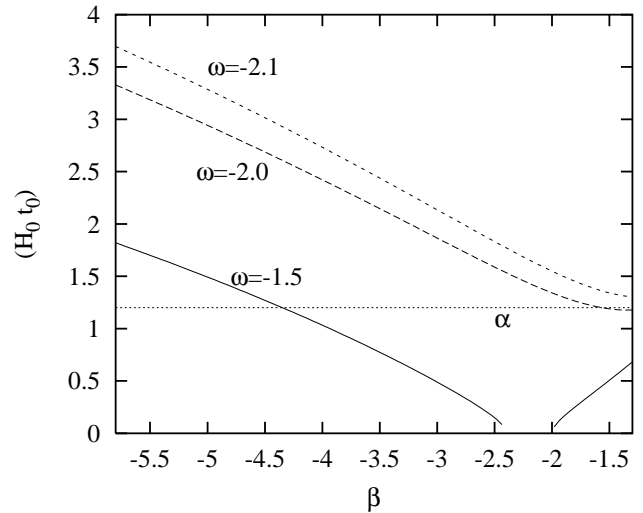}
\caption{$H_0 t_0$ {\it vs} $\beta$ for different $\omega$ values and $\alpha=1.2$}
\label{figure5}
\end{figure}

In figure 4 and 5 we plot $H_0t_0$ with respect to $\beta$ for $\alpha=1.1
~~{\rm and}~~1.2$. For this we consider different values of $\omega$.
For admissible values for the parameters the curves representing the 
expression $H_0t_0$ should intersect the horizontal line representing
a constant value of $\alpha$ 
at least once within the allowed range of $\beta$. 
It is interesting to find that for both the graphs the curves intersect the 
straightline representing $\alpha$ at least once when $\omega$ lies within 
the range $-2.0\leq\omega\leq-1.5$. 
in a different 
investigation Banerjee and Pavon\cite{Rban} have also arrived at the 
same range 
of $\omega$ 
while looking for late time acceleration in BD theory without a 
potential.
 In fact, in other investigations\cite{Rsen,Rban} 
in scalar tensor theories 
studying late time acceleration similar conclusions are made. 
In all these analyses the constraint coming from 
the solar system measurements
($\omega>600$) is not satisfied. We would discuss about this point in the 
discussion section.

%%%%%%%%%%%%%%%%%%%%%%%%%SECTION III
%%%%%%%%%%%%%%%%%%%%%%%%%%%%%%%%%%%%%%

\section{Density Perturbation}
While the accelerated expansion is supported by the
Brans Dicke
theory
with an appropriate form of the potential , it
is worth  checking whether
the issue of structure formation is modified by the
dynamics of the
Brans Dicke field treated here. We are particularly
interested in the
evolution of the density perturbation in the context of
Brans Dicke theory with the self interacting
potential. Our objective here is to
find the perturbations of the field equations (5)-(8)
and analyze the
 behaviour of the relevant variables
 for the accelerated expansion of the universe.
In reference
\cite{R23} a similar treatment is done to
analyze the behaviour of
the perturbed solutions in the original Brans
Dicke theory.
\par
As our interest is in cosmological density
perturbations, we will be using the
temporal component of the field equations.
Let us denote by $\delta g_{\mu \nu}$, $\delta R_{00}$
and $\delta T^{00}$ perturbations in $g_{\mu \nu}$,
$R_{00}$ and $T^{00}$, respectively. Denoting  $\delta
g_{\mu \nu}$ by $h_{\mu \nu}$,
$\delta R_{00}$ can be expressed as
\begin{equation}
\delta R_{00}={1\over{2R^2}}\left[\ddot h_{kk}-2{\dot
R\over{R}}\dot
h_{kk}+2\left({\dot R^2\over{R^2}}-{\ddot
R\over{R}}\right)h_{kk}\right]
\end{equation}

Similarly the perturbations in the trace of the
energy-momentum tensor
can be expressed as,
\begin{equation}
T=T+\delta T=T+\delta\rho-3 \delta p
\end{equation}
The perturbation of the d'Alembertian of the Brans
Dicke field is given
by
\begin{equation}
\delta\Box\phi=\delta \ddot\phi+R\dot R
h^{kk}\dot{\phi}-{1\over{2R^2}}\dot h_{kk}\dot\phi+3{\dot
R\over{R}}\delta
\dot\phi-{\nabla^2\over{R^2}}\delta\phi
\end{equation}
The following parameters are used for the relevant
perturbations
\begin{eqnarray}
h_{kk} &=& R^2h\nonumber \\
\delta\phi &=&
\lambda\phi~~~~~~{\rm{where}}~~~\lambda<<1\\
\delta\rho &=&
\Delta\rho~~~~~~{\rm{where}}~~~\Delta<<1\nonumber
\end{eqnarray}
where $h(t)~~,\lambda(t)~~{\rm{and}}~~\Delta(t)$ are
the perturbed
gravitational field, scalar field and matter energy
density respectively.
\par
As was mentioned in the earlier section, only three
of the four
field
equations are independent. Hence, three
equations are used for perturbation.
They are
\begin{eqnarray}
-3{\ddot R\over{R}} & = &
{\rho\over{\phi}}~{3\gamma_B(1+\omega)+2+\omega\over{2\omega+3}}\nonumber\\
&~~~~~ +
&{\bar{V_c}\over{2}}\left[1+{2-m\over{2\omega+3}}\right]\phi^{m-1}
+ {\ddot\phi\over{\phi}} +
\omega{\dot\phi^2\over{\phi^2}}
\end{eqnarray}
\begin{equation}
\Box\phi =
{\rho(1-3\gamma_B)\over{2\omega+3}}-{2-m\over{2\omega+3}}{V_c}\phi^m
\end{equation}
\begin{equation}
\dot\rho+3(1+\gamma_B){\dot R\over{R}}\rho=0
\end{equation}
where
\begin{eqnarray}
\gamma_B &=& {2-\beta\over{3\alpha}}-1\\
{\rm{and}}~~~m &=& {\beta-2\over{\beta}}
\end{eqnarray}
Using the above parameters, the set of
perturbed equations take the form
\begin{eqnarray}
{\ddot h\over{2}}+{\dot R\over{R}}\dot h & = &
{\rho\over{\phi}}(\Delta-\lambda)
~{3\gamma_B(1+\omega)+2+\omega\over{2\omega+3}}
\nonumber\\
&~~~~ - &
{\bar{V_c}\over{2}}\left[1+{2-m\over{2\omega+3}}\right](m-1)\lambda\phi^{m-1}\nonumber\\
&~~~~ + &
\ddot\lambda+2(\omega+1){\dot\phi\over{\phi}}\lambda
\end{eqnarray}
\begin{eqnarray}
\ddot \lambda & + & \dot
\lambda\left(2{\dot\phi\over{\phi}}+3{\dot
R\over{R}}\right)+\lambda\left({\ddot
\phi\over{\phi}}+3 {\dot
R\over{R}}{\dot\phi\over{\phi}}\right)-{\dot
h\over{2}}{\dot\phi\over{\phi}}-{\nabla^2\lambda\over{R^2}}\nonumber\\
&~~~ =
&{\Delta\rho(1-3\gamma_B)\over{\phi(2\omega+3)}} -
{2-m\over{2\omega+3}}\bar{V_c}\lambda m\phi^{m-1}
\end{eqnarray}
\begin{equation}
\dot\Delta-(\gamma_B+1)\left({\dot h\over{2}}-\delta
v^k_{,k}\right)=0
\end{equation}
where $v_\mu$ is the comoving velocity of the fluid.
The growth of perturbations primarily occur in the
post-recombination era. The pressure acting on the
matter is negligible during this era and hence,
it is appropriate to consider the matter to be
pressureless  i.e,  $\gamma_B=0$.  From (43) and (44),
$\alpha$ and $\beta$ can be expressed in terms
of $m$
\begin{eqnarray}
\beta &=& {2\over{1-m}}\\
{\rm{and}}~~~\alpha &=& -{2m\over{3(1-m)}}
\end{eqnarray}
Hence, the perturbed equations for the pressureless
case are
\begin{eqnarray}
{\ddot h\over{2}}+{\dot R\over{R}}\dot h & = &
{\rho\over{\phi}}(\Delta-\lambda)~{2+\omega\over{2\omega+3}}\nonumber\\
& - &
{\bar{V_c}\over{2}}\left[1+{2-m\over{2\omega+3}}\right](m-1)\lambda\phi^{m-1}\nonumber\\
& + &
\ddot\lambda+2(\omega+1){\dot\phi\over{\phi}}\lambda
\end{eqnarray}
\begin{eqnarray}
\ddot \lambda+\dot \lambda\left(2{\dot\phi\over{\phi}}
+ 3{\dot
R\over{R}}\right)+\lambda\left({\ddot
\phi\over{\phi}}+3 {\dot
R\over{R}}{\dot\phi\over{\phi}}\right) - {\dot
h\over{2}}{\dot\phi\over{\phi}}-{\nabla^2\lambda\over{R^2}}\nonumber\\
={\Delta\rho\over{(2\omega+3)\phi}}-{2-m\over{2\omega+3}}\bar{V_c}\lambda
m\phi^{m-1}
\end{eqnarray}
\begin{equation}
\dot\Delta-\left({\dot h\over{2}}-\delta
v^k_{,k}\right)=0
\end{equation}
The  perturbation in the four-velocity $\delta
v^k$ is to be set null, which
can be done by one infinitesimal gauge transformation.
Let us suppose
that the perturbation in the scalar field behave
as plane waves
\begin{equation}
\lambda(\vec{x}, t)=\lambda(t)\exp(-i\vec{k}.\vec{x})
\end{equation}
where $\vec{k}$ is the wave number of perturbation.
Using equation (52) and substituting the solutions
(9), (12) and (19)
in equation (50) and (51) we get
\begin{eqnarray}
\ddot\Delta & - &
{4m\over{3(1-m)}}{\dot\Delta\over{t}}-\ddot\lambda -
{4(1+\omega)\over{1-m}}{\dot\lambda\over{t}}=
~{\bar{\rho_c}\over{t^2}}(\Delta-\lambda)
~{2+\omega\over{2\omega+3}}\nonumber\\
 & - &
{\bar{V_c}\over{2}}\left[1-{m-2\over{2\omega+3}}\right](m-1){\lambda\phi_0^{m-1}t_0^2\over{t^2}}
\end{eqnarray}
\begin{eqnarray}
\ddot\lambda & = &
-{\dot\lambda\over{t}}{4-2m\over{1-m}}-{\lambda\over{t^2}}{2\over{1-m}}
+
{\dot\Delta\over{t}}{2\over{1-m}}+{k^2\lambda\over{{\cal{R}}_0^2}}\left({t\over{t_0}}\right)^{4m\over{3(1-m)}}\nonumber\\
& + & {\Delta\bar{\rho_c}\over{(2\omega+3)t^2}} -
{2-m\over{2\omega+3}}\bar{V_c}{\lambda
m\phi_0^{m-1}t_0^2\over{t^2}}
\end{eqnarray}
where
\begin{equation}
{\bar{\rho_c}}={\rho_c\over{\phi_0}}~t_0^{2\over{1-m}}\\
\end{equation}
Combining these two equations and neglecting the
higher order terms
beyond $t^{-2}$ (as the modes are analyzed in the
asymptotic region) we
get,
\begin{equation}
\ddot\Delta
+{C_1\over{t}}(\dot\Delta-\dot\lambda)+{C_2\over{t^2}}(\Delta-\lambda)+{C_3\over{t}}\dot\lambda+{C_4\over{t^2}}\lambda=0
\end{equation}
where
\begin{eqnarray}
C_1 &=& -{4m+6\over{3(1-m)}}\nonumber\\
C_2 &=&
-{\omega+3\over{2\omega+3}}\{{4\omega\over{(1-m)^2}}-{4(m+3)\over{3(1-m)^2}}-{4m-6\over{3(1-m)}}\}\nonumber\\
C_3 &=& -{10m+12\omega+6\over{3(1-m)}}\nonumber\\
C_4 &=&
{2\over{1-m}}+\{{m-1\over{2}}-{(m-2)(3m-1)\over{2(2\omega+3)}}\}\nonumber\\
&\{& {4(\omega+3+m)\over{3(1-m)^2}}\}-
{1\over{2\omega+3}}\{{4\omega\over{(1-m)^2}}\nonumber\\
&-&
{4(m+3)\over{3(1-m)^2}}-{4m-6\over{3(1-m)}}\}\nonumber
\end{eqnarray}
In order to solve the above equation we assume the
following form
\begin{equation}
\Delta-\lambda=f(t)~;~~~~f(t)=\xi
t^\delta~;~~~~\Delta=\chi t^\theta
\end{equation}
where $\xi$ and $\chi$ are constants.\\
With the above substitution we find that $\theta =
\delta$ and
\begin{equation}
\chi\theta^2+\theta[\chi(C_3-1)+\xi(C_1-C_3)]+C_4(\chi-\xi)+C_2\xi=0
\end{equation}

The solution for $\theta$ from the above equation
\begin{eqnarray}
& \theta_{\pm} & =
{1\over{2\chi}}[(\chi(1-C_3)+\xi(C_3-C_1))\nonumber\\
 &\pm&
\sqrt{\{\chi(1-C_3)+\xi(C_3-C_1)\}^2-4\chi\{C_4(\chi-\xi)+C_2\xi\}}]
\label{perturbation_modes}
\nonumber\\
\end{eqnarray}
Bertolami et al \cite{R16} have considered 
the growing modes for density perturbation in the asymptotic 
limit of $\vert \omega \vert>>1$. The
solution of $\theta$ in this limit for our case is
\begin{eqnarray}
& \theta_{\pm}& \rightarrow
{2\omega\over{1-m}}\{\left(1-{\xi\over{\chi}}\right)\nonumber\\
& \pm
&\sqrt{\left(1-{\xi\over{\chi}}\right)^2+{1-m\over{2\omega}}\left[1+{m\over{1-m}}{\xi\over{\chi}}\right]}\}
\end{eqnarray}
This asymptotic value of $\theta$ for large
$\vert\omega\vert$ depends
only on the power of the potential $m$. Since
$\beta$ is always
negative, m($={\beta-2\over{\beta}}$) is a positive
number. (Note that $m={\beta-2\over{\beta}} \neq 1$ as both $\alpha$ and $\beta$ blow up for $m=1$.) So, $\theta_+$
represents the growing mode for $\xi<\chi$.
Clearly for $m=2$ 
the mode matches with the result given by
Bertolami et
al\cite{R16}.
However, we found that $\omega$ 
should lie in the range $-1.5$ and $-2.0$. So in this connection it is 
worthwhile  
 to check if growing modes exist with this limitation.
For $\omega=-1.8$
we can calculate $\theta_{\pm}$ 
from the equation (\ref{perturbation_modes}). We find that with this value of 
$\omega$ also $\theta_+$ represents a growing 
mode. We, of course, need to make suitable choice for the values of 
$\xi$, $\chi$ and $m$ to achieve this. 
Thus in our case 
perturbations can grow with time for range of $\vert \omega\vert$ allowed 
in this model.

%%%%%%%%%%%%%%%%%SECTION IV%%%%%%%%%%%%%%%%%%%%%%

%%%%%%%%%%%%%%%%%%%%%%%%%%%%%%%%%%%%%%%%%%%%%%%%

\section{Discussion and Conclusions}
In this work we have investigated the nature of
potential relevant to
the power law expansion of the universe in a self-
interacting Brans
Dicke Cosmology with a perfect fluid distribution.
The age of
the universe and the
time variability of the gravitational coupling are also calculated.
The value of the parameters are constrained from
different physical conditions. We have graphically represented the
permissible values of the parameters.
We have also studied the evolution of density perturbations
in this model.

From our analysis we draw the following conclusions:
\begin{enumerate}
\item Accelerated growth of the scale factor can be
driven by this
positive power law potential with $\gamma_\phi$
lying in the range $-0.47\geq\gamma_\phi\geq -0.87$.
This range for $\gamma_\phi$ is very well consistent with the observations.
\item The gravitational coupling grows with time which agrees quite
well
with the observational facts\cite{R25}.
\item This model also allows growing modes
for the energy density
perturbation of matter implying that the
dynamics of the self interacting Brans Dicke field
does not upset the
structure formation scenario.
\end{enumerate}
This model depends entirely on three parameters $\alpha$, $\beta$ and 
$\omega$. The parameter $\alpha$ is only constrained to be greater than
$1$ by the fact that the universe is accelerating. Depending on the 
value of $\alpha$, the value of $\beta$ gets restricted within a certain range. 
The value of $\omega$ depends on these two parameters.
In order to ensure the weak 
(positivity) energy conditions as well as to produce the observed 
values of $\Omega_m$, 
$\alpha$ and $\gamma_\phi$, 
we require $\omega$ to be in the range 
$-2\leq\omega\leq-1.5$. Although this value of $\omega$ 
does not satisfy the solar system bound of 
$(\omega>600)$ this is a generic difficulty
in most of the investigations done 
in the context of scalar tensor theories \cite{Rsen,Rban}. 
In a completely different work 
in BD theory without potential Banerjee and Pavon\cite{Rban}
also had a similar 
range for $\omega$. 
The work of Bertolami and Martins\cite{R16}
is an exception where they
have obtained accelerated solution with large 
$|\omega|$. They have, however, not considered the positive energy condition 
for the energy density of the scalar field. 
A large value of $|\omega|$ necessitates, violation of 
this condition in these models.
The violation of the positive energy condition is particularly serious because the 
contribution of the energy density from matter is subdominant with 
respect to the missing energy density as is suggested by observation. 
 On the other hand 
this is not the only case when the solar system limit is violated. There
are other evidences in literature where small value of $|\omega|$ has 
been supported.The extended inflationary model suggested by La and 
Steinhardt\cite{LS} requires $\omega$ to be 20. 
In order to explain the structure 
formation scenario successfully in scalar tensor theory, the constraint 
on $\omega$ is not also compatible with the solar system test\cite{Rgaz}.\\ 
  
Aother imporant point to note is that as the universe expands in a constant 
power law fashion in this model, it has an acceleration also in the 
radiation dominated era which upsets the primeval nucleosynthesis scenario.
This problem also occurs if the universe is considered to have a power law 
expansion and no compromise can be done between the bigbang 
nucleosynthesis and present accelerated phase in general relativity 
as is shown by Kaplinghat et al\cite{Rkapling}.
Infact the existing supernova observational data provides sufficient 
evidence for the fact that the universe was decelerating in the near
past\cite{Rturner}.  
In a recent work of Banerjee and Pavon\cite{Rban} 
it has been shown that in BD theory the nucleosynthesis 
problem can be avoided by 
considering $\omega$ to be a function of $\phi$. But 
in that case also $\omega$ asymptotically acquires a small negative 
value to have a late time accelerating phase.\\   

In the end we can say that though the model presented here describes 
the present day universe quite successfully, it disagrees at two other 
important point of observation. While the age of the universe, density 
parameters and $\gamma_\phi$  calculated here satisfies the observed 
constraints, the power law accelerated expansion of the universe upsets 
the standard decelrating phase of the universe until recent past
 and the important solar system 
limit on $\omega$. So to get rid of these two problem one approach is 
to consider $\omega$ a function of $\phi$ so that local inhomogeneities 
give rise to large $\omega$ and one could get both accelerating and 
decelerating phase. We wish to address this problem  
in our next work.

%%%%%%%%%%%%%%%%%%%%%%%%%%%%%%%%%%%%%%%%%%%%%%%%%%%%%%%%

\section{Acknowledgement}
We would like to thank Indrajit Chakrabarty for helping us to
incorporate the figures in the paper and Anjan Ananda Sen for 
useful comments.

\end{document}